\documentclass[conference, anonymous]{IEEEtran}
\IEEEoverridecommandlockouts
\usepackage{cite}
\usepackage{amsmath,amssymb,amsfonts}
\usepackage{algorithmic}
\usepackage{float}
\usepackage{graphicx}
\usepackage{subcaption}
\usepackage{textcomp}
\usepackage{xcolor}
\usepackage{booktabs}  % For better table lines
\usepackage{tabularx}  % For flexible table width
\usepackage{multirow}  % For merging cells vertically
\def\BibTeX{{\rm B\kern-.05em{\sc i\kern-.025em b}\kern-.08em
    T\kern-.1667em\lower.7ex\hbox{E}\kern-.125emX}}
\usepackage{tikz}
\usetikzlibrary{positioning, shapes.geometric,arrows.meta}

\begin{document}

\title{Hierarchical Attention Network for Interpretable ECG-based Heart Disease Classification}

\author{
\IEEEauthorblockN{Mario Padilla Rodriguez}  
\IEEEauthorblockA{\textit{Engineering and Computer Science} \\  
University of Detroit Mercy \\
Detroit, MI, USA\\  
padillma1@udmercy.edu}  
\and  
\IEEEauthorblockN{Mohamed Nafea}  
\IEEEauthorblockA{\textit{Electrical and Computer Engineering} \\  
Missouri University of Science and Technology \\
Rolla, MO, USA\\  
mnafea@mst.edu}  

\thanks{*Work in progress.} \thanks{
A preliminary component of this study was conducted by Mario Padilla Rodriguez as part of his Master's thesis in Electrical and Computer Engineering at the University of Detroit Mercy.}}
\maketitle

\begin{abstract}
Cardiovascular disease remains one of the leading causes of mortality worldwide, underscoring the need for accurate as well as interpretable diagnostic machine learning tools. In this work, we investigate heart disease classification using electrocardiogram (ECG) data from two widely-utilized datasets: The MIT-BIH Arrhythmia and the PTB-XL datasets.
%In this work, we explore heart disease classification using electrocardiogram (ECG) data based on two widely-used datasets: The MIT-BIH Arrhythmia and the PTB-XL datasets. 
We adapt a hierarchical attention network (HAN), originally developed for text classification, into an ECG-based heart-disease classification task. Our adapted HAN incorporates two attention layers that focus on ECG data segments of varying sizes. We conduct a comparative analysis between our adapted HAN and a more sophisticated state-of-the-art architecture, featuring a network with convolution, attention, and transformer layers (CAT-Net). Our empirical evaluation encompasses multiple aspects including test accuracy (quantified by 0-1 loss); model complexity (measured by the number of model parameters); and interpretability (through attention map visualization). Our adapted HAN demonstrates comparable test accuracy with significant reductions in model complexity and enhanced interpretability analysis: For the MIT-BIH dataset, our adapted HAN achieves 98.55\% test accuracy compared to 99.14\% for CAT-Net, while reducing the number of model parameters by a factor of 15.6. For the PTB-XL dataset, our adapted HAN achieves a 19.3-fold reduction in model complexity compared to CAT-Net, with only a 5\% lower test accuracy. From an interpretability perspective, the significantly simpler architecture and the hierarchical nature of our adapted HAN model facilitate a more straightforward interpretability analysis based on visualizing attention weights. Building on this advantage, we conduct an interpretability analysis of our HAN that highlights the regions of the ECG signal most relevant to the model's decisions.
\end{abstract}


\section{Introduction}
A recent state-of-the-art model for ECG-based heart disease classification, termed as CAT-Net, combines convolutional layers, attention mechanisms, and transformers \cite{islam2024cat}. Designed for single-lead ECG data, CAT-Net achieves a 99.14\% test accuracy on the MIT-BIH arrhythmia dataset \cite{moody2001impact}. Despite its high accuracy, CAT-Net's high complexity and computational demands make its results challenging for medical practitioners to interpret. The {\it{primary contribution of this work is an adaptation of the hierarchical attention network (HAN) \cite{yang2016hierarchical}, originally developed for document/text classification, for ECG-based heart disease classification}}. We empirically demonstrate that {\it{by exploiting the hierarchical structure of the ECG data segments, our adapted HAN model achieves comparable test accuracy to CAT-Net, while significantly reducing model complexity}}. Furthermore, our adapted HAN model {\it{facilitates easier interpretability due to its simplicity and hierarchical nature, allowing visualization of certain regions of the ECG signal that are most relevant to the model's decisions}}. This work benchmarks our adapted HAN against CAT-Net, comparing their performance (test accuracy), model complexity (in terms of parameter count), and attention-based interpretability.

The rest of this paper is organized as follows. Sections \ref{related} \& \ref{prelims} discuss related work and preliminaries. Section \ref{method} presents the methodologies we utilize. Section \ref{eval} presents our experimental evaluation. Section \ref{Con} concludes the paper.

\section{Related work} 
\label{related}
\cite{islam2024cat} introduced CAT-Net, a single-lead ECG-based network that combines convolutional, attention, and transformer layers. Convolutional layers are used to extract local features while ``multi-head" attention layers \cite{vaswani2017attention} capture contextual information. To address class imbalance, CAT-Net applies adapted versions of the synthetic minority oversampling technique (SMOTE) \cite{nitesh2002smote,batista2004study}. It achieves a test accuracy of 99.14\% on the MIT-BIH arrhythmia dataset \cite{moody2001impact}, making it the highest-performing single-lead ECG classification model.
%to generate additional minority class samples
\cite{abubaker2022detection} developed a deep, 38-layer, convolutional neural network (CNN) for cardiovascular disease (CVD) detection using images of 12-lead ECGs, rather than actual time-series data, achieving  98.23\% test accuracy. Similarly, [\citenum{khan2021cardiac}] proposed the pre-trained MobileNet-V2 architecture which uses images of ECG signals, scoring 98\% test accuracy. \cite{tan2018application} developed a stacked CNN-Long Short Term Memory (LSTM) model for coronary artery disease (CAD) detection, achieving 98.55\% test accuracy. The authors used three different ECG databases to generate 64,000 modified ECG segments. %with added noise. 

\cite{mousavi2020han} proposed utilizing a hierarchical attention network (HAN)--introduced in \cite{yang2016hierarchical} for text classification-- for the {\it{binary classification task}} of detecting atrial fibrillation from 12-lead ECG signals, achieving a 98.81\% accuracy on the MIT-BIH dataset. The HAN architecture mirrors the hierarchical structure of documents as it processes words (tokens) first and then sentences composed of these words. At each level, an encoder and an attention layer identify the most {\it{relevant}} content. Similarly, the HAN-ECG model in \cite{mousavi2020han} treats the ECG signal as a hierarchy, with waves at the lowest level (see Figure \ref{fig:ECG}), heartbeats composed of multiple waves and windows of several beats. Each level employs its own encoder and attention mechanism to highlight crucial local features. Compared to \cite{mousavi2020han}, our proposed model remains closer to the original HAN architecture in \cite{yang2016hierarchical} by dissecting the ECG signal into fixed-length sequences, each consisting of fixed-size segments. Further, we consider a {\it{multi-class heart-disease classification task}}. Empirical evidence via hyperparameter tuning demonstrates that, for the multi-class task, considering two levels of hierarchy for the ECG signal leads to superior results compared to three levels as in \cite{mousavi2020han} (See Section \ref{perf-results}). 

Several works incorporate interpretability analysis in heart-disease detection. \cite{ayano2022interpretable} presents an overview of interpretable ECG-based methods. \cite{anand2022explainable} uses the SHAP framework from \cite{lundberg2017unified} to highlight segments of the ECG signal most relevant for their model's predictions. Another work, \cite{goodfellow2018towards}, uses class activation maps (CAM) \cite{zhou2016learning} to visualize the regions of an ECG signal for which a CNN model gives the highest attention weights.

In natural language processing (NLP), a prevalent interpretability technique is to visualize attention weights \cite{wang2016attention,lin2017structured,ghaeini2018interpreting,serrano2019attention}. The attention mechanism  \cite{bahdanau2014neural} generates probability distributions over the input, which are widely considered as indicators of feature importance. This approach can be applied to identify specific regions of an input ECG signal crucial for prediction. Indeed, \cite{mousavi2020han} and \cite{hong2019mina} implement multiple attention layers and visualize their weights to determine the most attended waves and beats in the signal. Nevertheless, despite its popularity, the direct explainability of attention has been questioned by prior research \cite{jain2019attention,wiegreffe2019attention,serrano2019attention}. Consequently, a separate line of work has focused on devising enhanced explanation methods for attention models \cite{sundararajan2017axiomatic,kobayashi2020attention,hao2021self,Meister2021sparse,liu2022rethinking}. In this work, we focus on visualizing attention weights to highlight critical regions of the ECG signal for model predictions, prioritizing simplicity of exposition and effectively demonstrating the superiority of our proposed model.

\section{Preliminaries} \label{prelims}
\subsection{Electrocardiogram Data}
\begin{figure}
    \centering
\includegraphics[width=0.45\textwidth]{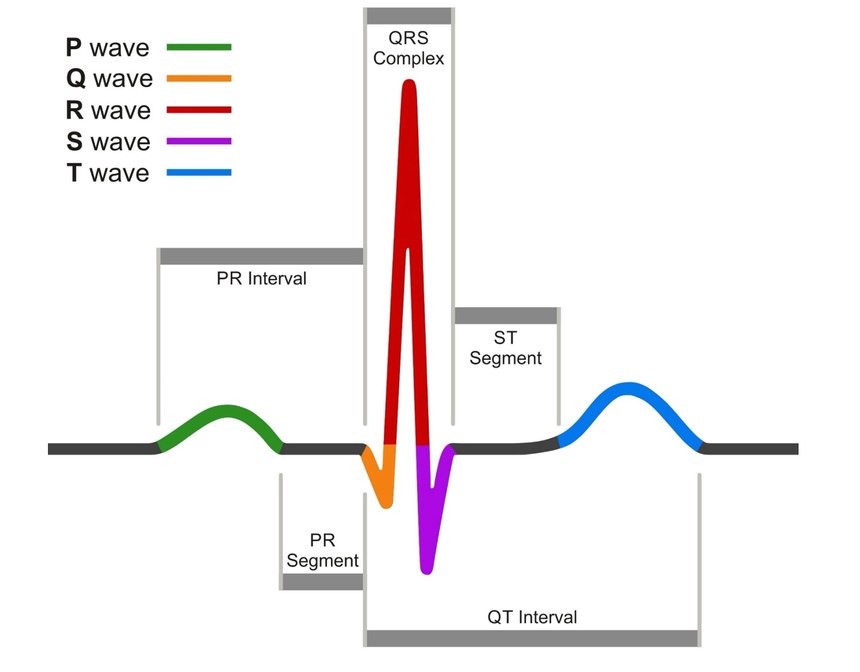}
    \caption{Components of an ECG Signal}
    \label{fig:ECG}
\end{figure}

Electrocardiogram (ECG) data is vital in cardiology, providing detailed information about the heart's electrical activity. Generated by electrodes placed on the skin, ECG recordings measure electrical impulses triggering heartbeats, allowing assessment of cardiac rhythm, rate, and other parameters. The resulting waveform reflects various cardiac cycle phases, including P wave, QRS complex, and T wave, as demonstrated in Figure \ref{fig:ECG}. ECG data is crucial for diagnosing and monitoring cardiovascular conditions, with its continuous nature enabling detection of transient and chronic abnormalities not evident in a single snapshot.

Specific ECG signal behaviors can indicate various heart diseases. Irregularities in intervals between characteristic waves, such as prolonged or shortened QT intervals, may signal conditions like long QT syndrome \cite{longQT2022}. Abnormal P-R intervals can indicate atrioventricular block \cite{cadogan2021PR}. Changes in waveform morphology, such as ST-segment elevation or depression, can reflect conditions like myocardial infarction or ischemia \cite{kashou2023ST}. Abnormal T waves may indicate heart failure or electrolyte imbalances. Arrhythmias are commonly detected through ECG analysis. Atrial fibrillation is characterized by erratic electrical signals in the atria, leading to irregularly spaced R-R interval and distinct absence of discernible P waves \cite{burns2023atrial}. Ventricular tachycardia is marked by rapid, wide, and abnormal QRS complexes \cite{lome2024ventricular}. Recognizing these and other ECG signal abnormalities is crucial for accurate diagnosis of heart diseases. Advanced ML models are increasingly used to identify these patterns automatically, improving diagnostic efficiency and enabling timely intervention.

\subsection{Interpretability of deep models}
Attention mechanisms are a widely employed technique for enhancing interpretability in deep learning models, particularly in NLP tasks such as document classification. Specifically, attention weights are computed and visualized to identify which words in a text sequence are most critical for the model's decision-making. 
However, while attention weights offer plausible explanations \cite{wiegreffe2019attention}, their reliability as faithful explanations has been questioned \cite{serrano2019attention,jain2019attention,bai2021attentions,Meister2021sparse,liu2022rethinking}. Several studies have evaluated the robustness of attention as a justification for model reasoning, with some proposing alternatives to the exclusive use of unprocessed attention weights. For example, \cite{serrano2019attention} examines the reliability of unprocessed attention weights for explainability by conducting ablation studies, removing features with the highest attention scores and observing the impact on model decisions.
Their empirical findings across various datasets reveal that in certain instances, the removal of highly-attended features does not significantly alter the model's output. As an alternative, \cite{serrano2019attention} proposes a feature importance ranking method that combines attention weights with their corresponding input gradients, ensuring that critical features possess both high attention scores and substantial gradients. However, this approach does not guarantee an entirely reliable ranking system. Another alternative methodology, presented in \cite{abnar2020quantifying}, leverages graph theory concepts to more accurately trace the propagation of attention weights through the model.

\section{Methodology}
\label{method}
We conduct a comparative analysis of two models for ECG-based heart disease classification. The first model is an adapted version of the hierarchical attention network (HAN); initially developed for document classification tasks in \cite{yang2016hierarchical}. The second model is a recently introduced state-of-the-art architecture called CAT-Net (convolution-attention-transformer network), specifically designed for ECG-based heart disease classification \cite{islam2024cat}. Through empirical evaluation on two distinct benchmark datasets, namely the MIT-BIH arrhythmia and the PTB-XL datasets, we demonstrate that our adapted HAN, despite its significantly reduced computational complexity and enhanced interpretability, achieves classification accuracy comparable to that of the more sophisticated CAT-Net architecture.

\subsection{Hierarchical attention network (HAN)}
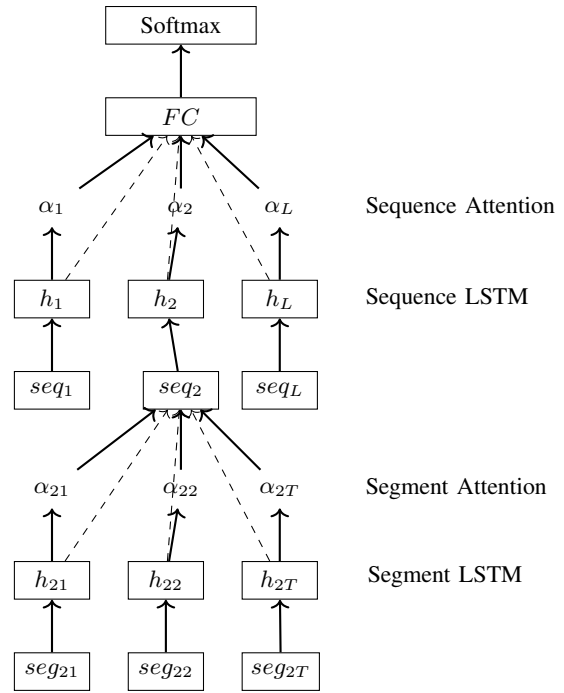
\begin{figure}
\centering
\begin{tikzpicture}[
    node distance=1.5cm,
    every node/.style={draw, minimum width=1cm, minimum height=0.5cm, align=center},
    arrow/.style={->, thick},
    dashedarrow/.style={->, dashed, thin},
    font=\small
]

\node (seg21) {${seg_{21}}$};
\node[right=0.5cm of seg21] (seg22) {${seg_{22}}$};
\node[right=0.5cm of seg22] (seg2T) {${seg_{2T}}$};

\node[above=0.7cm of seg21] (h21) {${h_{21}}$}; 
\node[right=0.5cm of h21] (h22) {${h_{22}}$};
\node[right=0.5cm of h22] (h2T) {${h_{2T}}$};
\draw[arrow] (seg21) -- (h21);
\draw[arrow] (seg22) -- (h22);
\draw[arrow] (seg2T) -- (h2T);
\node[above=0.7cm of h21, draw=none] (alpha21) {${\alpha_{21}}$};
\node[right=0.7cm of alpha21, draw=none] (alpha22) {${\alpha_{22}}$};
\node[right=0.3cm of alpha22, draw=none] (alpha2T) {${\alpha_{2T}}$};
\draw[arrow] (h21) -- (alpha21);
\draw[arrow] (h22) -- (alpha22);
\draw[arrow] (h2T) -- (alpha2T);
\node[above=0.8cm of alpha21] (seq1) {${seq_1}$}; % Sequence representation
\node[above=0.8cm  of alpha22] (seq2) {${seq_2}$};
\node[above=0.8cm of alpha2T] (seqL) {${seq_L}$}; % Final sequence representation
\draw[arrow] (alpha21) -- (seq2);
\draw[arrow] (alpha22) -- (seq2);
\draw[arrow] (alpha2T) -- (seq2);
\draw[dashedarrow] (h21) -- (seq2);
\draw[dashedarrow] (h22) -- (seq2);
\draw[dashedarrow] (h2T) -- (seq2);
\node[above=0.7cm of seq1] (h1) {${h_1}$}; % LSTM for ECG sequences
\node[right=0.5cm of h1] (h2) {${h_2}$};
\node[right=0.5cm of h2] (hL) {${h_L}$};
\draw[arrow] (seq1) -- (h1);
\draw[arrow] (seq2) -- (h2);
\draw[arrow] (seqL) -- (hL);
% ECG Sequence Attention Layer (no boxes for alphas)
\node[above=0.7cm of h1, draw=none] (alpha1) {${\alpha_1}$};
\node[right=0.7cm of alpha1, draw=none] (alpha2) {${\alpha_2}$};
\node[right=0.3cm of alpha2, draw=none] (alphaL) {${\alpha_L}$};
\draw[arrow] (h1) -- (alpha1);
\draw[arrow] (h2) -- (alpha2);
\draw[arrow] (hL) -- (alphaL);
\node[above=0.7cm of alpha2, draw, minimum width=2cm] (v) {$FC$}; % Weighted sum output
\draw[arrow] (alpha1) -- (v);
\draw[arrow] (alpha2) -- (v);
\draw[arrow] (alphaL) -- (v);
% Add new arrows from h_21, h_22, h_2T to seq_2
\draw[dashedarrow] (h1) -- (v);
\draw[dashedarrow] (h2) -- (v);
\draw[dashedarrow] (hL) -- (v);
% Final Softmax Layer (at the top)
\node[above=0.7cm of v, draw, minimum width=2cm] (softmax) {Softmax};
\draw[arrow] (v) -- (softmax);
% Annotations on the right
\node[above right=0.7cm and 0.3cm of seg2T, align=left, draw=none] (labels) {
    \begin{tabular}{l}
    Segment LSTM \\
    \end{tabular}
};

\node[right=0.3cm of alpha2T, align=left, draw=none] (labels) {
    \begin{tabular}{l}
    Segment Attention \\
    \end{tabular}
};

\node[right=0.3cm of hL, align=left, draw=none] (labels) {
    \begin{tabular}{l}
    Sequence LSTM \\
    \end{tabular}
};

\node[right=0.3cm of alphaL, align=left, draw=none] (labels) {
    \begin{tabular}{l}
    Sequence Attention \\
    \end{tabular}
};
\end{tikzpicture}
\caption{Hierarchical Attention Network (HAN)}
\label{fig:HAN}
\end{figure}

The architecture of our adapted HAN, shown in Figure \ref{fig:HAN}, closely mirrors the original architecture for document classification in \cite{yang2016hierarchical}. It consists of a convolutional layer, followed by an ECG segment encoder, a segment-level attention layer, a sequence encoder, and a sequence-level attention layer. The original architecture in \cite{yang2016hierarchical} used gated recurrent units (GRUs), which we replace in this work by long short-term memory (LSTM) blocks to process the time-series ECG data. 

The model first processes small segments of the signal consisting of 30 samples each, then examines sequences  composed of 10 segments, totaling 300 samples per input. This allows the HAN to first identify important segments (akin to words) and subsequently focus on larger sequences (akin to sentences). By analyzing segments, the model learns local patterns and by examining sequences, it learns the relationship among individual segments, capturing long-range dependencies. 

Attention layers are used to highlight the regions of the ECG signal that are most critical for model's prediction, both at the segment and sequence levels. Let $W$ and $b$ denote the parameters of the attention layer. For an input $X=[X_1,\cdots,X_n]$, the layer's attention weights $\alpha$ and output are computed as:
\begin{align}
    &v = \tanh(X W + b), \quad \alpha = \frac{\exp(v)}{\sum_i \exp(v_i)} \\
    & {\rm{Output}} = \sum_i\alpha_{i} X_{i}
\end{align}
The model is completed with a fully-connected (FC) layer and dropout regularization to reduce overfitting. The training is optimized using Adam's algorithm \cite{kingma2014adam}. With the exception of \cite{mousavi2020han,hong2019mina},  existing work fails to leverage the inherent hierarchical structure of ECG signals. Compared to \cite{mousavi2020han,hong2019mina}, our model is composed of a two-level HAN (as in \cite{yang2016hierarchical}) instead of three levels. Further, our model is designed for a {\it{multi-class heart-disease classification task}}, as opposed to the binary classification tasks considered in \cite{mousavi2020han,hong2019mina}. We empirically demonstrate that two levels of hierarchy provide superior performance for ECG-based multi-class heart-disease classification. We also benchmark our HAN against the state-of-the-art CAT-Net, described in the next section.

\subsection{Convolution-attention-transformer network (CAT-Net)}
CAT-Net \cite{islam2024cat} is comprised of a sequence of {\it{four}} convolutional and attention layers for feature extraction, followed by a transformer encoder with {\it{multi-head attention}} to capture long-term signal dependencies. Let $h$ denote the number of attention heads, and $Q, K, V$ denote the query, key, and value matrices, respectively.  $W^Q, W^K, W^V$ are the learned parameter matrices for each head, and
$W^O$ is the output projection matrix. Multi-head attention is computed as: 
\begin{align}
&\text{MultiHead}(Q, K, V) = \text{Concat}(\text{head}_1, \ldots, \text{head}_h) W^O \\
& \text{head}_i = \text{Attention}(Q W^Q_i, K W^K_i, V W^V_i)
\end{align}
In \cite{islam2024cat}, CAT-Net was trained on the MIT-BIH Arrhythmia and INCART \cite{tihonenko2008st} datasets. 

\subsection{Datasets}
Unlike for electronic health records (EHR), there exist several publicly available ECG datasets for scientific research. We evaluate our adapted HAN and the CAT-Net on two publicly available ECG datasets, namely the MIT-BIH Arrhythmia Database \cite{moody2001impact} and the PTB-XL dataset \cite{wagner2020ptb}.

\subsubsection{MIT-BIH database}\label{MIT-BIH}
This database contains 48 recordings, each 30 minutes long, of two-channel (2-lead) ambulatory ECGs. The recordings were collected and analyzed by Boston's Beth Israel Hospital (BIH) and the Massachusetts Institute of Technology (MIT) between 1975 and 1979. In most recordings, one channel represents a modified limb lead II (MLII), obtained by placing electrodes on the chest, while the other channel is typically lead V1, also placed on the chest. The MLII lead provides a view of the heart’s electrical activity that looks from the right arm to the left leg, while the V1 lead provides a view of the heart’s right side. 

For this study, 23 recordings were randomly selected from a pool of 4000 24-hour ambulatory ECG recordings collected from a mixed population of inpatients (about 60\%) and outpatients (about 40\%) at Beth Israel Hospital. The remaining 25 recordings were chosen from the same pool to ensure the inclusion of less common but clinically significant arrhythmias that might be underrepresented in a smaller random sample. The dataset includes recordings from 25 male subjects, aged 32 to 89 years, and 22 female subjects, aged 23 to 89 years.

The recorded signals were originally analog, but were converted to digital. They were first filtered to prevent analog-to-digital converter (ADC) saturation and to perform anti-aliasing. This filtering used a passband filter of 0.1 to 100 Hz. The filtered signals were then digitized at a rate of 360 Hz per signal, with 11-bit resolution, using hardware developed at the MIT Biomedical Engineering Center and the BIH Biomedical Engineering Laboratory. Each recording contains about 648,000 samples. 

In addition to the raw signal data, each recording includes an annotations file with labels for every heartbeat, marked by the R-peak. The labs at MIT-BIH used a slope-sensitive QRS detector to generate a preliminary set of beat labels. Each 30-minute recording was then reviewed independently by two cardiologists, to refine these labels by adding any missed beats, removing false detections, and correcting labels for abnormal beats. Since its introduction in 1989, only seventeen beat labels have needed correction, demonstrating the reliability of the annotations for locating each beat. Therefore, for this dataset, we do not use an algorithm to detect beats, we rather directly access the annotation files for the 48 recordings.

The MIT-BIH dataset includes 15 types of heartbeats which are divided into five different classes: Normal (N), Supraventricular ectopic beat (S), Ventricular ectopic beat (V), Fusion beat (F), and Unknown beat (Q). N represents regular heartbeats with no abnormalities, indicating a normal sinus rhythm. S refers to abnormal heartbeats originating above the ventricles, often premature and irregular. V consists of premature heartbeats originating from the ventricles, which can lead to more serious arrhythmias. F results from the simultaneous activation of both the atria and ventricles, creating a hybrid complex. Q includes heartbeats that do not fit into the other categories, often due to noise or unclassifiable patterns. 

\subsubsection{PTB-XL dataset}\label{PTB-XL}
This dataset is significantly larger than the MIT-BIH database, containing 21,837 clinical 12-lead ECGs from 18,885 patients, each 10 seconds long. The recordings were collected between October 1989 and June 1996. Each ECG was analyzed and labeled by two independent cardiologists, and a single recording could include multiple statements—following the SCP-ECG standard \cite{bousseljot1995nutzung}—regarding rhythm, diagnosis, and structure. Although not used in this study, each recording also includes demographic information about the patient, such as age and gender.

Of the 18,885 patients, 52\% are male and 48\% are female, with ages ranging from 0 to 95 years. The dataset contains 71 diagnostic classes, which have been grouped into the following five superclasses (with the number of records in each class in parentheses): NORM (9,528), normal ECG; MI (5,486), myocardial infarction; STTC (5,250), ST/T change; CD (4,907), conduction disturbance; and HYP (2,655), hypertrophy.

For each of the two datasets described above, {\it{our adapted HAN model is designed to classify each ECG signal to one of the corresponding five classes.}}  
\subsection{Data pre-processing} \label{data-pre-process}
For a fair comparison, we adopt the same ECG signal pre-processing steps for the two models. In particular, we use the same data preprocessing techniques in \cite{islam2024cat}. These include denoising the signal using the discrete wavelet transform (DWT) which decomposes the signal into different frequency components, each with a resolution that matches its scale. DWT provides both time and frequency information, making it particularly useful for analyzing non-stationary signals like ECG. More specifically, the signal is first passed through a series of high-pass and low-pass filters, resulting in a high-frequency component (called detailed component) and a low-frequency component (called approximation component). This process is repeated to further decompose the approximation component into multiple levels. The high-frequency components, which often contain noise, are processed by applying a threshold, where coefficients below this threshold are set to zero, effectively removing noise while preserving important signal features. Finally, the signal is reconstructed by combining the processed high-frequency components and the approximation component. This step reverses the decomposition process to obtain a denoised version of the original signal. 

The next step is to locate all the R-peaks in each ECG recording. For the MIT-BIH dataset, this step is straightforward since each recording comes with an annotation file containing the locations of each R-Peak, reviewed by two cardiologists. For the PTB-XL dataset (and most other existing ECG datasets), a peak detection algorithm is required to locate the R-peaks. We tried several existing algorithms including the Pan-Tompkins \cite{pan1985real}, Hamilton \cite{hamilton1986quantitative}, and WQRS \cite{zong2003robust}; with the Pan-Tompkins algorithm achieving the highest performance.

The ECG signal is then segmented using the locations of the R-peaks. For the MIT-BIH dataset, a window size of 300 samples is used, where 99 samples are taken before (left of) the R-peak, and 201 samples are taken after (right of) the R-peak. For the PTB-XL, a window size of 350 samples is used to better accommodate any errors made by the peak detection algorithm, where 150 samples are taken before the R-peak, and 200 samples are taken after the R-peak. A 60\%--20\%--20\% training-validation-testing split is used for both datasets. 

For the MIT-BIH data (Section \ref{MIT-BIH}), since an overwhelming majority of samples belong to the ``Normal" class, the SMOTE algorithm \cite{nitesh2002smote} is applied to generate more samples from the minority classes, and the majority class is under-sampled. Note that the SMOTE algorithm is only applied to the training data. To ensure robust generalization, oversampling of the minority classes is done while maintaining the original ratios found in the dataset. For the majority class undersampling, 50,000 out of 72,000 samples are randomly selected. This approach differs slightly from the SMOTE method used with CAT-Net \cite{islam2024cat}, where each of the minority classes is oversampled to exactly 50,000 samples. No class balancing was applied to the PTB-XL (Section \ref{PTB-XL}) dataset since all of its five classes are equally represented.

Additionally, while the signals in the MIT-BIH dataset are 2-lead and in the PTB-XL dataset are 12-lead, we only utilize one lead for both models: MLII for MIT-BIH and V5 for PTB-XL. A model built for single-lead ECG data is not only more interpretable, as it avoids the complexity of multiple leads (some of which may not always be useful), but also more practical, since a medical practitioner may not always have access to all leads.

\subsection{Interpretability}
In addition to comparing its performance with the state-of-the-art CAT-Net model, we conduct an interpretability analysis for our adapted HAN. We do so by visualizing the segment-level and sequence-level attention weights, as discussed in Section \ref{related}. Compared to CAT-Net, the HAN provides a more direct attention-based interpretation where the inputs to each attention level correspond to predefined segments of the ECG-signal. This allows for identifying the regions of the ECG signal that are most critical for predictions, see Section \ref{Int-results}. 

\section{Experimental results} \label{eval}
In this section, we present our empirical evaluation for the two datasets presented in the previous section using both our adapted HAN and the CAT-Net learning models. 

\subsection{Hyperparameter tuning}
To obtain the best performance for each of the datasets and models under investigation, we tuned the hyperparameters of our models as follows. 

\subsubsection{MIT-BIH dataset} For the CAT-Net, we utilized the exact same tuned hyperparameters in \cite{islam2024cat} that led to the highest accuracy performance on the MIT-BIH dataset, namely, a learning rate of $0.01$; $128$ fully connected (FC) hidden units (neurons) for the FC layer; and $0.2$ dropout rate for regularization. 

For our adapted HAN, we tuned its hyperparameters as follows. We tried a learning rate in the range $[0.001-0.1]$; a stride for the convolutional layer of $1$ and $2$; a number of units for the LSTM layers in the range $[64-256]$; a number of neurons for the FC layer in the range $[64-256]$, and a dropout rate in the range $[0.1-0.5]$. For the convolutional layer, we utilized the same number of filters and filter-size as in \cite{yang2016hierarchical}, namely $16$ filters, each of size $21$. 

\subsubsection{PTB-XL dataset} For this dataset, we tuned the hyperparameters of both our adapted HAN and the CAT-Net models, since this dataset was not studied in \cite{islam2024cat}. For both models, we tried a learning rate ranging from $0.0005$ to $0.01$; a number of neurons for the FC layers ranging from $64$ to $512$; and a dropout rate ranging from $0.1$ to $0.5$. For our HAN model, additional tuning was performed on (i) the number of convolutional filters, ranging from $32$ to $128$ filters; (ii) the size of each convultional filter, ranging from $5$ to $25$; and (iii) the number of LSTM units, ranging from $32$ to $64$.  

\subsection{Performance results} \label{perf-results}
\subsubsection{MIT-BIH dataset} Table \ref{tab:ecg_model_combined} demonstrates train and test accuracies for our adapted HAN and the CAT-Net models, trained on the MIT-BIH dataset. As mentioned in the experimental setup, the same data pre-processing steps are used for both models, and both are trained for 30 epochs. Given the numerous hyperparameter configurations we tried, only the best-performing configuration for each model is reported. For our HAN model, the best performing hyperparameter configuration is a learning rate of $0.001$; a $128$ FC neurons; $16$ convolutional filters each of size $21$; $64$ units in both LSTM layers, and a $0.2$ dropout rate.

\begin{table}[h!]
\centering
\caption{Model accuracies for the MIT-BIH dataset}
\begin{tabularx}{\columnwidth}{|l|X|X|X|}
\hline
\textbf{Model} & \textbf{Train Accuracy (\%)} & \textbf{Test Accuracy (\%)} & \textbf{\# Parameters} \\
\hline
HAN & $99.01$ & $98.55$ & $76,265$ \\
\hline 
CAT-Net & $99.98$ & $99.14$ & $1,189,637$ \\
\hline
\end{tabularx}
\label{tab:ecg_model_combined}
\end{table}

The results in Table \ref{tab:ecg_model_combined} demonstrate that our adapted HAN model achieves comparable accuracy performance to the much deeper and more complex CAT-Net, yet with a significant ($15.6$ fold) reduction in the number of model parameters. The main purpose of this study is to demonstrate that ``{\emph{exploiting the hierarchical structure of the ECG signal could lead to drastic saving in the complexity of the learning model, for the same level of target accuracy}}''. Further, our HAN model is more amenable for interpretability analysis since (i) our model architecture is drastically simpler than that of the CAT-Net and, as mentioned previously, (ii) the inputs to the hierarchical attention layers of our model correspond to pre-defined regions of the ECG signal, hence permits visualizing the regions in the ECG signal that are most relevant to the model's decisions (See Section \ref{Int-results}).

\subsubsection{PTB-XL dataset}
Similarly, the performance of both models on the PTB-XL dataset is summarized in Table \ref{tab:ptb_results}; reporting the best-performing hyperparameter configuration for each model. For the HAN, the best configuration is a learning rate of $0.005$; a $256$ neurons for the FC layers; a dropout rate of $0.3$, $64$ convolutional layer filters, each of size $9$, and $64$ units in the two LSTM layers. The best performance achieved by the CAT-Net uses a learning rate of $0.001$, $256$ FC neurons, and a dropout rate of $0.2$. 

\begin{table}[h!]
\centering
\caption{Model accuracy on the PTB-XL dataset}
\begin{tabularx}{\columnwidth}{|l|X|X|X|}
\hline
\textbf{Model} & \textbf{Train Accuracy (\%)} & \textbf{Test Accuracy (\%)} & \textbf{\# Parameters} \\
\hline
HAN & $80.60$ & $75.59$ & $94,157$ \\
\hline
CAT-Net & $89.65$ & $80.11$ & $1,812,997$ \\
\hline
\end{tabularx}
\label{tab:ptb_results}
\end{table}

The highest test accuracies achieved by our HAN model and the CAT-Net are $75.59\%$ and $80.11\%$, respectively. Our adapted HAN still achieves closer performance to CAT-Net (with about $5\%$ lower accuracy), yet with a significant reduction in number of parameters ($19.3$ fold). We note that the difference between training and test accuracies for the CAT-Net is approximately $10\%$, while for our HAN model, it is around $5\%$. Tuning the dropout rate  does not significantly affect regularization for either model. 

As an attempt to improve performance on the PTB-XL dataset, we trained a modified version of our adapted HAN model with three LSTM and three attention layers (i.e., three levels of hierarchy instead of two). We performed an extensive hyperparameter tuning procedure and obtained highest training and testing accuracies of $79.25\%$ and $74.58\%$, respectively. This performance however is worse than the best results we obtained with two level of hierarchy (cf. Table \ref{tab:ptb_results}), while requiring more model parameters, particularly, $144,661$ compared to $94,157$ with two hierarchy levels. Therefore, we observe that considering two levels of hierarchy for ECG signals results in a superior performance compared to that obtained with three levels as in \cite{mousavi2020han}. Finally, we note that a potential reason for the degraded performance of both models on the PTB-XL dataset could be due to the inaccuracies of the utilized Pan-Tompkins peak detection algorithm (cf. Section \ref{data-pre-process}). We leave the exploration of the reasons behind the high-variance of both models, especially the for CAT-Net, on the PTB-XL dataset, for future work. 

\subsection{Interpretability results} \label{Int-results}
\begin{figure}[htbp]
    \centering
    \begin{subfigure}[b]{\linewidth}
        \centering
        \includegraphics[width=\linewidth]{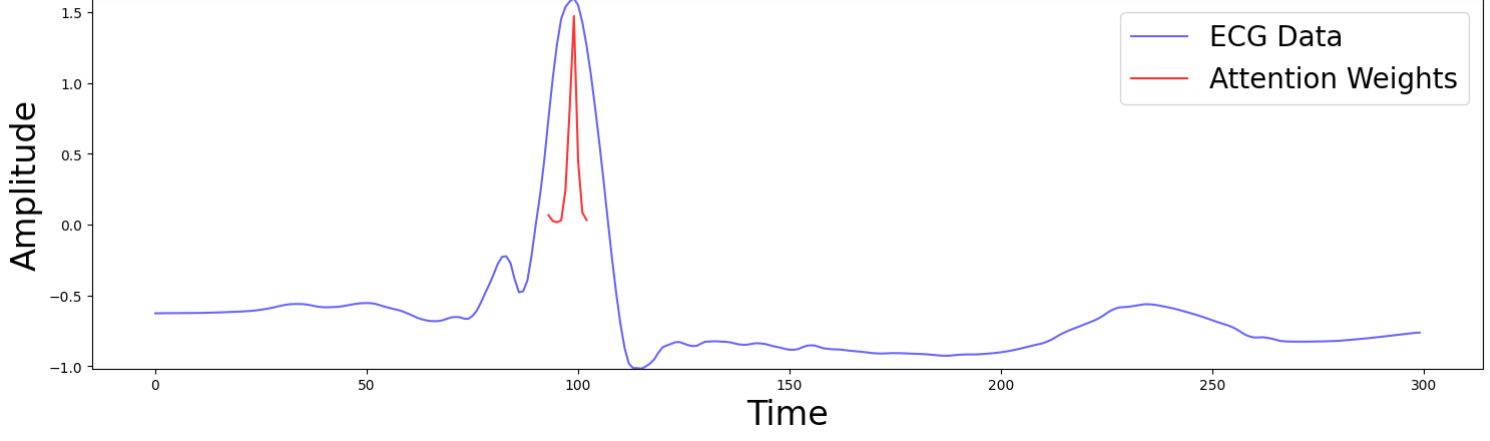}
        \caption{Segment attention weights of a random sample selected from the MIT-BIH dataset.}
        \label{fig:seg_att_weights}
    \end{subfigure}
    
    \begin{subfigure}[b]{\linewidth}
        \centering
        \includegraphics[width=\linewidth]{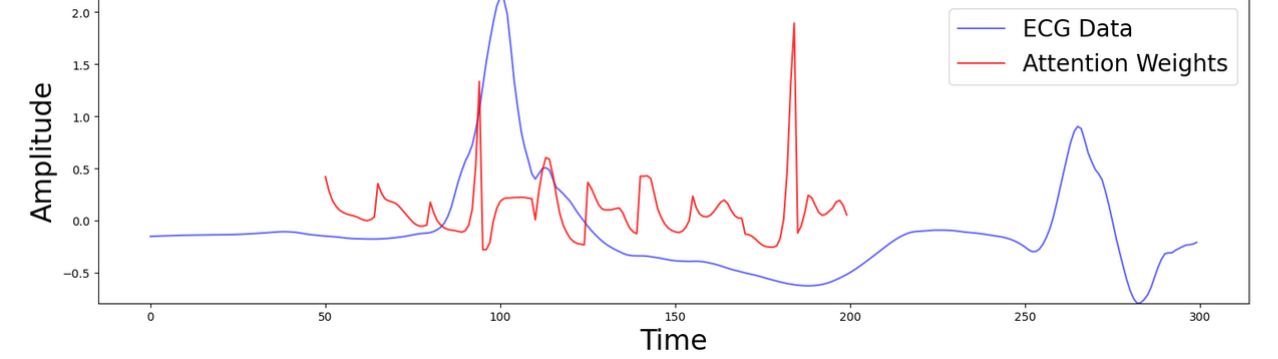}
        \caption{Sequence attention weights of a random sample selected from the MIT-BIH dataset.}
        \label{fig:seq_att_weights}
    \end{subfigure}
    
    \caption{Attention weights visualized for segment-level and sequence-level attention for two randomly selected samples from the MIT-BIH dataset.}
    \label{fig:combined_attention_weights}
\end{figure}

In this section, we visualize the attention weights for our HAN model (trained on the MIT-BIH dataset) against their corresponding ECG segments/sequences, to identify the regions of the ECG signal that are most influential on the model's decisions. Recall that each sequence of the ECG signal consists of $10$ consecutive segments. Figure \ref{fig:combined_attention_weights} displays the segment-level and sequence-level attention weights for two different randomly selected signals from the MIT-BIH dataset. 

In Figure \ref{fig:seg_att_weights}, the segment-level attention weights are shown for the region characterized by the QRS complex, specifically the R-peak, which is detected using the Pan-Tompkins algorithm (cf. Section \ref{data-pre-process}). The segment-level attention weights exhibit their highest value around the R-peak; hence this visualization highlights the importance of the R-peak for heart-disease identification, which aligns with medical knowledge. 

Figure \ref{fig:seq_att_weights} shows the sequence-level attention weights across the overall ECG signal, demonstrating that some sequences of the ECG signal are more influential on the model's decision than others. They also highlight the importance of the T-wave, which is another physiologically meaningful feature of the ECG signal. Note that, for simplicity of exposition and to focus on the main contributions of this work, we only capture and visualize attention weights. Implementing methods that further process attention weights for more faithful/reliable explanations, as in \cite{sundararajan2017axiomatic,kobayashi2020attention,hao2021self,Meister2021sparse,liu2022rethinking,serrano2019attention,abnar2020quantifying}, is left for future work.

\section{Conclusion and Discussion}\label{Con}
This work demonstrates the efficacy of our adapted hierarchical attention network (HAN) for ECG-based heart disease classification, achieving near state-of-the-art performance with significantly reduced model complexity and a higher level of interpretability. Our results for the MIT-BIH dataset in particular show that our adapted HAN model achieves a comparable test accuracy to that of the significantly more complex convolution-attention-transformer network (CAT-Net) while reducing the number of model parameters by a factor of 15.6. Furthermore, the simplicity of our adapted HAN  model and its hierarchical architecture permits a more straightforward interpretability analysis based on visualizing attention weights for fixed-length segments, and sequences of segments, of the ECG signal. This highlights key regions of the ECG signal relevant for diagnosis made by the machine learning model. 

{\it{Importantly, our results establish a valuable insight to the similarity between the tasks of document classification and ECG-based heart-disease classification, where for both tasks, it is evident that exploiting the hierarchical structure of the data could potentially lead to improved performance.}} 

In future work, we plan to explore alternative interpretability methods beyond visulaizing standalone (unprocessed) attention weights. Additionally, we plan to further investigate the reasons behind the lower performance and high variance of both studied models on the PTB-XL dataset. 

\bibliographystyle{IEEEtran}
\bibliography{IEEEabrv,mybibfile}

\end{document}